\documentclass[manuscript]{aastex}

\usepackage{amssymb}
\usepackage{longtable}
\usepackage{array}
\usepackage{txfonts}

\usepackage{amsmath}

\slugcomment{Not to appear in Nonlearned J., 45.}

\shorttitle{Solar Spatial Magnetic Field}
\shortauthors{Liu et al.}

\begin{document}


\title{A Statistical Study on Property of Spatial Magnetic Field for Solar Active Region}

%% WITH SOLAR DYNAMICS OBSERVATORY/ATMOSPHERIC IMAGING ASSEMBLY

\author{S. Liu\altaffilmark{1}}


\altaffiltext{1}{Key Laboratory of Solar Activity, National
Astronomical Observatories, Chinese Academy of Science, Beijing
100012, China}

\begin{abstract}
Magnetic fields dominate most solar activities, there exist direct relations
between solar flare and the distributions of magnetic field, and also its corresponding magnetic energy.
In this paper, the statistical results about the relationships between the spatial magnetic field and solar
flare are given basing on vector magnetic field observed by the Solar Magnetic Field Telescope (SMFT)
at Huairou Solar Observing Station (HSOS). The spatial magnetic fields are obtained by extrapolated
photosphere vector magnetic field observed by SMFT. There are 23 active regions with flare eruption
are chosen as data samples, which were observed from 1997 to 2007. The results are as follows:
1. Magnetic field lines become lower after flare for 16(69\%) active regions; 2. The free energy are decreased after
flare for 17 (74\%) active regions. It can conclude that for most active regions the changes of magnetic field after solar flare
are coincident with the previous observations and studies.

\end{abstract}




\section{INTRODUCTION}
Most solar activities such as filaments eruptions, flares and coronal
mass ejections (CMEs) are dominated by magnetic fields.
All these eruptions phenomena are energetic events due to explosive release of magnetic
field energy \citep{kra82, shi95, wan94,tsu96, for00,bon01, pri03, nin12}.
The magnetic field distributions and magnetic energies contained in active regions play the key roles in producing solar flares.
Hence, the knowledge of magnetic field is necessary to understand solar activities \citep{val05,wie12}.
Previous researchers have studied lots of magnetic parameters related to the productions of flares,
such as the magnetic flux of the active region \citep{mci90}, the shear angle \citep{lv93},
the inductive electric field \citep{liu08} and the magnetic helicity \citep{nin02, nin03}.
It is undoubted that there should exist changes of magnetic field before and after flares, since
the sources of kinetic and thermal energy of flare are the magnetic energy release.
The enhancements of transverse magnetic field were found after flare for some active regions \citep{wan05, su11, wan12},
which match the theoretical models \citep{hu08, hu11, li11, wan12}.
More precisely, it is a fact that the flare energy comes from magnetic free energy, which is magnetic energy available
in corona for conversion into kinetic and thermal energy \citep{pri07}.
So the studies of relationships between free energy and flares may reveal more physical essences of flare
productions. For example, \citet{jin09} found a positive correlation between the free energy and the flare index \citep{abr05},
\citet{sun12} found there are evident decreases of free energy after the solar flares.

The whole spaces above the active regions are filled with magnetic fields,
whose evolutions should be exhibited in whole spaces above active regions.
However the magnetic fields with sufficient resolution and accuracy usually be measured on the photosphere,
while magnetic field measurements in chromosphere and corona are only available for a few special cases \citep{sol03, lin04,van08}.
To study the properties of spatial magnetic fields, the magnetic field extrapolations should be applied actually.
Force free extrapolation is an classical method used to study solar spatial magnetic field recently.
Force free extrapolations are based on the assumption that there is no Lorentz force in the whole space of active region,
which can be expressed by $(\nabla\times\textbf{\emph{B}}) \times\textbf{\emph{B}}=0$ mathematically.
The spatial/coronal magnetic fields can be reconstructed from this physical
model (namely, force free model: $(\nabla\times\textbf{\emph{B}}) \times\textbf{\emph{B}}=0$,
$\nabla\cdot\textbf{\emph{B}}=0$), in which the observed photospheric
magnetic fields are taken as a boundary conditions
 \citep{wu90, mic94, ama97, sak81, yan00, whe00, wie04, son06, he08, liu11a, liu11b}.
 It means that coronal magnetic fields are considered to be force free, while at the boundary it is
connected to the photospheric magnetic fields observed.

In this paper, force free extrapolation \citep{son06, liu11b} is used to obtained spatial magnetic field of active region,
then the properties of magnetic field topology and free magnetic energy are
studied. The main researches are targeted at the changes of magnetic field topology and free magnetic energy
before and after solar flares, and the results are gained by large sample statistically.
The remainder of this paper is organized as
follows. In Section 2 the data and extrapolation method used are presented. In Section 3 the results are shown.
At last, the discussions and conclusions are given in Section 4.

\section{OBSERVATIONS DATA AND EXTRAPOLATION METHOD}

The photosphere magnetograms used as boundary for extrapolation are observed by Solar
Magnetic Field Telescope (SMFT) at Huairou Solar Observing Station (HSOS).
SMFT consists of 35 cm refractor with a vacuum tube, birefringent filter,
CCD camera including an image processing system \citep{ai86}.
The birefringent filter is tunable, working either at the photosphere line Fe I $\lambda$ 5324.19 \AA,
with a 0.150 \AA ~bandpass, or at the chromosphere line, H$\beta$, with a 0.125 \AA ~bandpass.
The line of Fe I $\lambda$ 5324.19 \AA ~(Lande factor $g$
=1.5) formed around the solar photosphere, is used for photospheric magnetic field observations. The
bandpass of the birefringent filter is about 0.15 \AA~for Fe I $\lambda$~5324.19
\AA~line. The center wavelength of the filter can normally be shifted -0.075 \AA~ relative to center of Fe I $\lambda$~
5324.19 \AA~ to measure of the longitudinal magnetic field and then the line center is applied to measure the transverse one \citep{ai86}.
Vector magnetograms are reconstructed from four narrow-band images of Stokes parameters ($I$, $Q$, $U$
and $V$). $V$ is the difference of the left and right circularly polarized images, $Q$ and $U$ are the differences between
two orthogonal linearly polarized images for different azimuthal directions, $I$ is the intensity derived from either the sum of two circularly polarized images is the line-of-sight field measurements or of two linearly polarized images in the transverse field measurement.
When $I$, $Q$, $U$ and $V$ are measured, the corresponding white light images are simultaneously obtained, which are employed to compensate for
the time differences during the measurements of $I$, $Q$, $U$ and $V$.
The sequence of obtaining Stokes images is as follows: First acquired the V/I image
; next the Q/I images, then the U/I image. The time required to
obtain a set of Stokes images was about 45 seconds. Each
image is associated with 256 integrated frames. To reconstruct the vector magnetograms, the linear relation is necessary between the magnetic field and the Stokes parameters $I$, $Q$, $U$
and $V$, which is true under the weak-field approximation \citep{jef89, jef91}:
\begin{equation}
B_{L}=C_{L}V ,\\
\end{equation}
\begin{equation}
B_{T}=C_{T}(Q^{2}+U^{2})^{1/4}, \\
\end{equation}
\begin{equation}
\theta=arctan(\dfrac{B_{L}}{B_{\bot}}),\\
\end{equation}
\begin{equation}
\phi=\dfrac{1}{2}arctan(\dfrac{U}{Q}),\\
\end{equation}
where $B_{L}$ and $B_{T}$ are the line-of-sight and transverse component of the photospheric field, respectively.
$\theta$ is the inclination between the vector magnetic field and the direction normal to the solar surface and $\phi$ is the field azimuth.
$C_{L}$ and $C_{T}$ are the calibration coefficients for the
longitudinal and transverse magnetic fields, respectively.
In this study, $C_{L}$ and $C_{T}$ are 8381 G and 6790 G \citep{wan96,su04},
respectively, which are obtained by theoretical calibration.
After routine data processing of HSOS data, the spatial
resolution of observational data is actually 2 arcsec/pixel $\times$
2 arcse/pixel and 3$\sigma$ noise levels of vector magnetograms are 20
G and 150 G for longitudinal and transverse components,
respectively.
To resolve $180^{\circ}$ ambiguity, the method of minimum energy are employed,
this method minimizes simultaneously both the divergence of the magnetic field ($\nabla\cdot\textbf{\emph{B}}$) and the electric current density ($\textbf{\emph{J}}=\nabla\times\textbf{\emph{B}}$) using a simulated annealing algorithm \citep{mat94,mat06}.
%\citep{wan94,wang97,wang01,mat06}.
At last, the components of the vector magnetic field
projected and re-mapped to local heliographic coordinates are used for magnetic field extrapolation.
%$(B_{x}=B_{T}cos\phi$, $B_{y}=B_{T}sin\phi$ ($B_{T}=\sqrt{B_{x}^{2}+B_{y}^{2}}$) and $B_{z}=B_{L}cos\theta$) are calculated in local heliocentric coordinates. %To minimize the projection effects, the requirement that the horizontal width of an active region is less than 30 degree is added for each magnetogram.

The approximate vertical integration (AVI) method
\citep{son06} was improved from the
direct integration method \citep{wu90} basing on equations (5)-(8).
\begin{equation}
 \dfrac{\partial B_{x}}{\partial z} = \dfrac{\partial B_{z}}{\partial x} + \alpha
 B_{y},
\end{equation}
\begin{equation}
 \dfrac{\partial B_{y}}{\partial z} = \dfrac{\partial B_{z}}{\partial y} - \alpha
 B_{x},
\end{equation}
\begin{equation}
 \dfrac{\partial B_{z}}{\partial z} = -\dfrac{\partial B_{x}}{\partial x} -
  \dfrac{\partial B_{y}}{\partial y},
\end{equation}
\begin{equation}
 \alpha B_{z} = \dfrac{\partial B_{y}}{\partial x} - \dfrac{\partial B_{x}}{\partial
 y}.
\end{equation}

The magnetic field is reconstructed by the following
formula firstly in AVI method,
\begin{equation}
\label{q-avi1}
\textbf{B}_{x} = \xi_{1}(x,y,z)F_{1}(x,y,z),
\end{equation}
\begin{equation}
\textbf{B}_{y} = \xi_{2}(x,y,z)F_{2}(x,y,z),
\end{equation}
\begin{equation}
\label{q-avi2}
\textbf{B}_{z} = \xi_{3}(x,y,z)F_{3}(x,y,z),
\end{equation}
assuming the second-order continuous partial
derivatives in a certain height range, 0$ <z<$H (H is the
height from the photosphere). In Equations
(\ref{q-avi1})-(\ref{q-avi2}), $\xi_{1}, \xi_{2}$ and $\xi_{3}$
mainly depend on $z$ and slowly vary with $x$ and $y$, while $F_{1},
F_{2}$ and $F_{3}$ mainly depend on $x$ and $y$ and weakly vary with
$z$, which are mathematical representation of the similarity solutions.
In solar active regions, we cannot seek analytical solutions for the magnetic field,
but we can construct an analytical asymptotic solutions within a thin layer.

After constructing the magnetic field, the following integration equations,
%(\ref{a})-(\ref{b})

%
\begin{equation}
\begin{array}{c}
\dfrac{d\xi_{1}}{dz}F_{1}(x_{i},y_{j},z) = \xi_{3}\dfrac{\partial F_{3}(x_{i},y_{j},z)}{\partial x}
+ \alpha(x_{i},y_{j},z)\xi_{2}F_{2}(x_{i},y_{j},z), \\[0.5cm]
\dfrac{d\xi_{2}}{dz}F_{2}(x_{i},y_{j},z) = \xi_{3}\dfrac{\partial F_{3}(x_{i},y_{j},z)}{\partial y}
- \alpha(x_{i},y_{j},z)\xi_{1}F_{1}(x_{i},y_{j},z),\\[0.5cm]
\dfrac{d\xi_{3}}{dz}F_{3}(x_{i},y_{j},z) = -\xi_{1}\dfrac{\partial F_{1}(x_{i},y_{j},z)}{\partial x}
 - \xi_{2}\dfrac{\partial F_{2}(x_{i},y_{j},z)}{\partial y}, \\[0.5cm]
\alpha(x_{i},y_{j},z)\xi_{3}F_{3}(x_{i},y_{j},z) = \xi_{2}\dfrac{\partial
F_{2}(x_{i},y_{j},z)}{\partial x}
 - \xi_{1}\dfrac{\partial F_{1}(x_{i},y_{j},z)}{\partial y}, \\[0.5cm]
{\rm for}   \hspace{3ex} 0    \leq z \leq \Delta z.
\end{array}
\end{equation}
are used to carry out the extrapolation. The solutions of the above equations can be found in \citet{son06}.
The performances of this AVI extrapolation method are evaluated in some previous studies.
Through the related investigations and comparisons, it is concluded that this method is acceptable for corona magnetic field
extrapolation \citep{son06, son07,liu11a,liu11b}. Also there are some actual applications about this AVI extrapolation to study the chromosphere and corona magnetic field \citep{son07,li07,liu11b}.

%The optimization method is used to obtain coronal magnetic field in this study.
%Optimization method proposed by \citet{whe00} and
%developed by \citet{wie04} consists in minimizing a joint
%measure for the normalized Lorentz force and the divergence of the
%field, given by the function,
%\begin{equation}
%\label{opit} L = \int_{V}\omega(x,y,z)[B^{-2}|(\nabla \times
%\textbf{B}\times \textbf{B}) |^{2}+|\nabla\cdot
%\textbf{B}|^{2}]d^{3}x,
%\end{equation}
%where $\omega(x,y,z)$ is a weighting function related position. It is clear that (for $w > 0$) the force-free equations are fulfilled
%when $L$ is equal to zero. This method involves minimizing $L$ by optimizing the solution function $\textbf{B}(x, t)$ through states
%that are increasingly force- and divergence-free, where $t$ is an artificial time-like parameter introduced.

\section{RESULTS}
In this work 23 active regions with flare eruptions observed by SMFT from 1997 to 2007 are
selected as data sample. The criteria of the selection for active regions are as follows: 1. The magnetic fields
were obtained at a time interval of $\sim$ 3 hours before and after solar flare, it means that there are individual observation
before and after flare, and the time differences between flare and observations should not be longer than 3 hours.
2. The solar flare should be a single one, it requires that there must be only one flare within the observations time interval of before and after
this flare, there should be no redundant flare occurring during this time interval.
3. The magnetic components of $B_{x}$, $B_{y}$ and $B_{z}$ can be reconstructed normally and completely, there should be no
observation data with defects are used in the work.
4. The longitudes of active regions observed and used for study should locate within 40 degree away from solar disk center.
5. The deviations of magnetic fluxes between positive and negative magnetic field should not exceed 20\%, which can be express
by formula $(F_{p}-F_{n})/F_{p}\le 20\%$ ($F_{p}$ and $F_{n}$ indicate positive and negative magnetic fluxes, respectively).
Through a rigorous selection, 23 active regions selected and listed in Table 1 are chosen for this study.
In the table, the number of active region (NOAA), the observation time, the flare eruption time,
the flare level and the deviation of flux are given.

\begin{figure}
\epsscale{1.0} \plotone{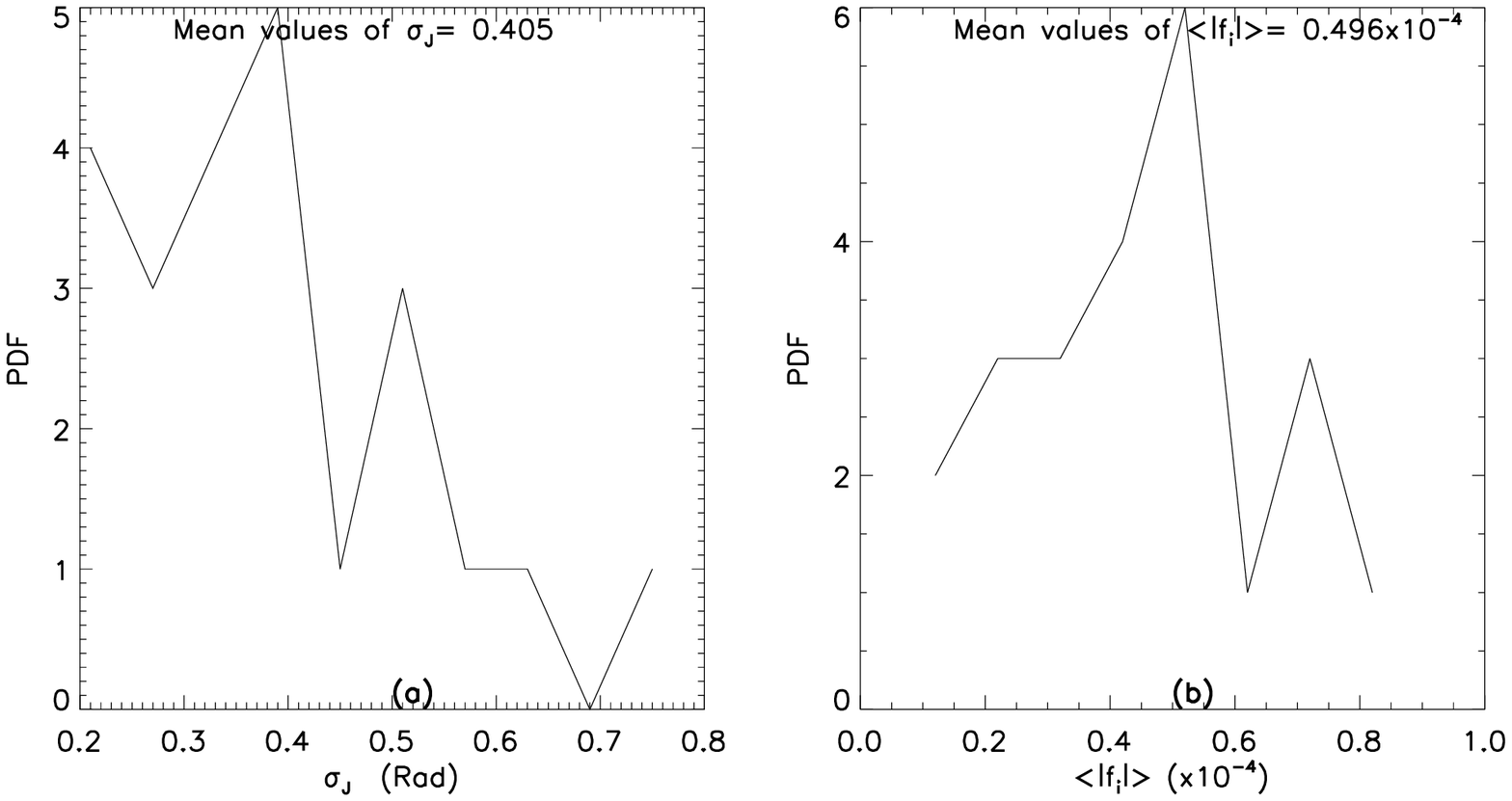} \caption{The distributions of $\sigma_{J}$ and $<|f_{i}|>$ for all active regions studied, and the corresponding mean values
are given in each plot individually.}\label{Fig01}
\end{figure}

\begin{figure}
\epsscale{1.0} \plotone{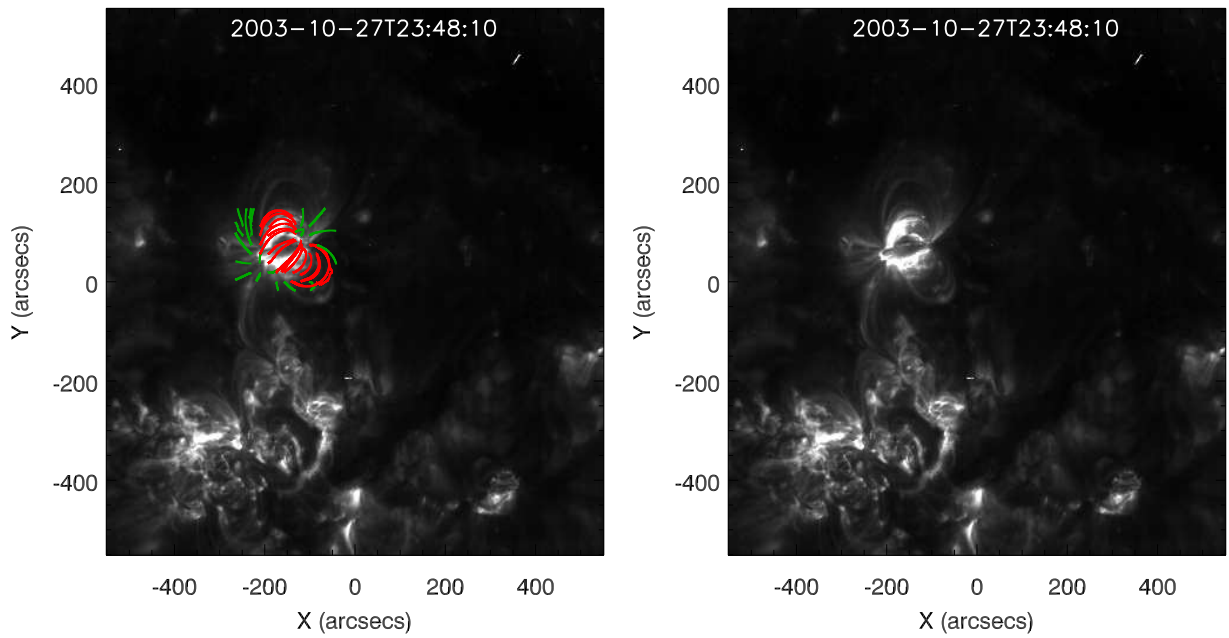} \caption{The left: the distributions of magnetic field lines are over plotted on the coronal images. The right: the coronal images obtained by EIT/MDI at 2003-10-27T23:48:10UT.}\label{Fig0-1}
\end{figure}

\begin{figure}
\epsscale{1.0} \plotone{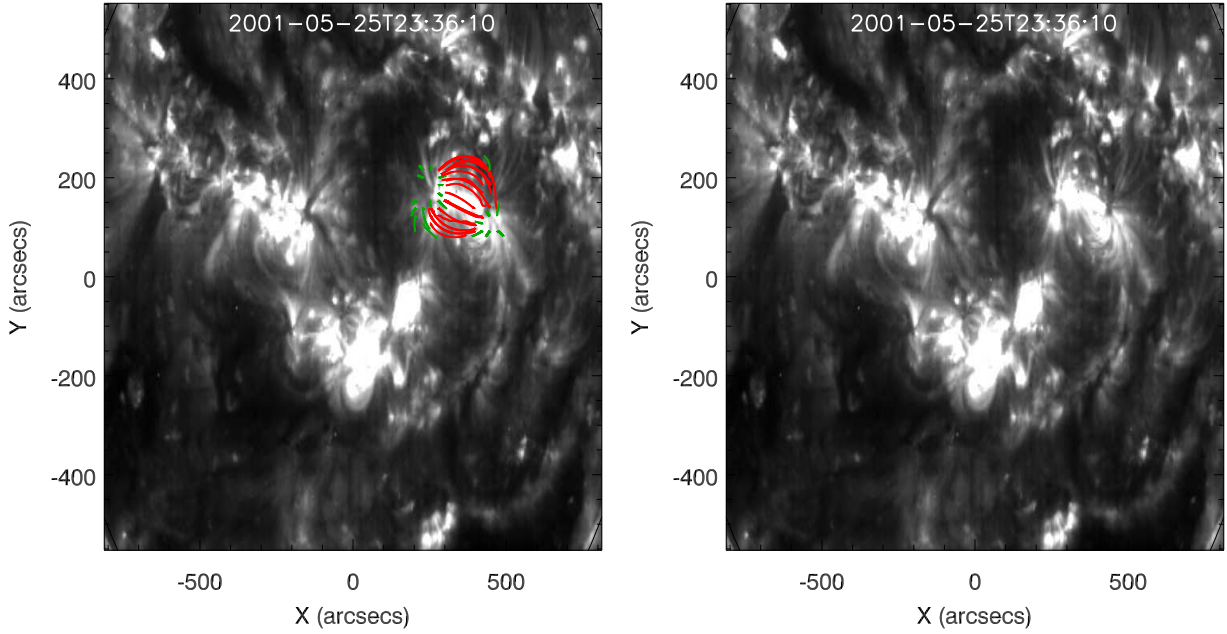} \caption{Same as Fig2, but the observed time is 2001-05-25T23:36:10UT.}\label{Fig0-2}
\end{figure}

\begin{figure}
\epsscale{1.0} \plotone{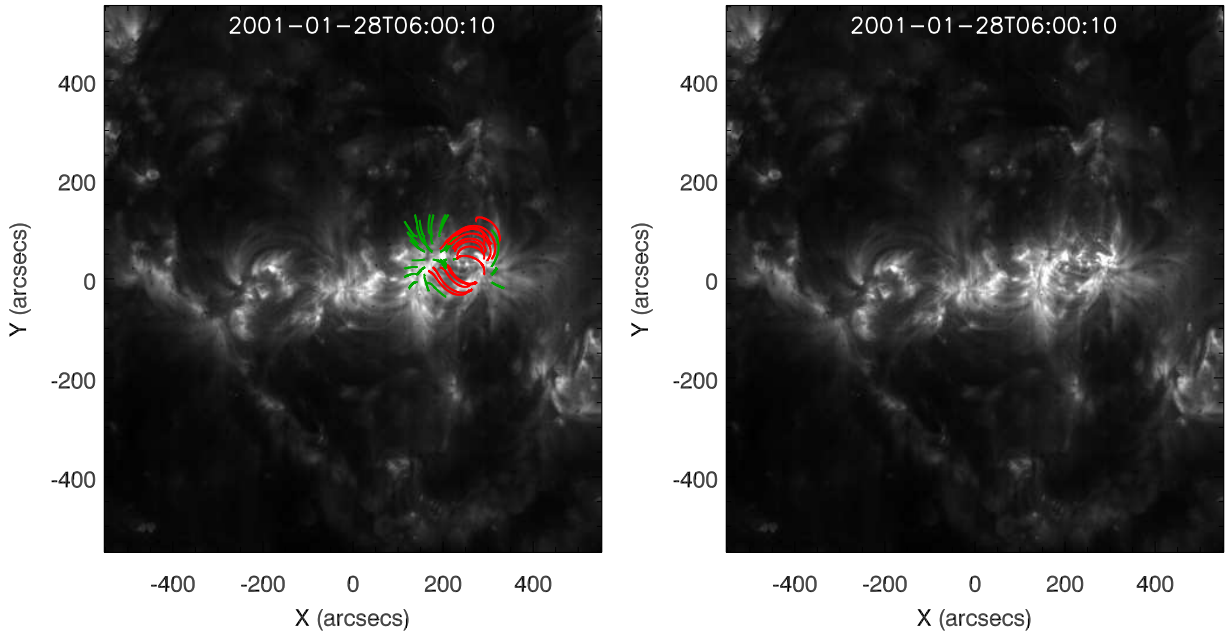} \caption{Same as Fig2, but the observed time is 2001-01-28T06:00:10UT.}\label{Fig0-3}
\end{figure}

The extrapolated fields should represent the solutions of force-free equations ($(\nabla\times\textbf{\emph{B}}) \times\textbf{\emph{B}}=0$,
$\nabla\cdot\textbf{\emph{B}}=0$) approximately, 
so the extent of force- and divergence-freeness of extrapolated field
should be checked at first.  The criterion of force-freeness $\sigma_{J}$,
\begin{equation}
\label{sigma_J}
 \sigma_{J} = \dfrac{\sum_{i}J_{i}\sigma_{i}}{\sum_{i}J_{i}},
\end{equation}
where
\begin{equation}
\sigma_{i} = {\rm sin} \theta_{i} = \dfrac{\mid \textbf{J} \times
\textbf{B} \mid_{i} }{J_{i}B_{i}},
\end{equation}
%.\ref{sigma_J}
and the criterion of divergence-freeness $f_{i}$,
%.\ref{f_i}
\begin{equation}
\label{f_i}
 f_{i} = \dfrac{{\int_{\bigtriangleup S_{i}} \textbf{B}
 \cdot \textbf{dS}}}{{\int_{\bigtriangleup S_{i}} \mid\textbf{B}
 \mid\textbf{dS}}} \thickapprox \dfrac{(\nabla\cdot \textbf{B})_{i}
 \bigtriangleup\textbf{V}_{i}}{B_{i}A_{i}},
\end{equation}
are used to check the quality of the extrapolated field \citep{whe00, der09}. $\sigma_{J}$ are
the weighted average of the sine of angle between the magnetic field and current
density, and it is related to Lorentz force. %($\textbf{F}\sim \textbf{J}\times \textbf{B}$). 
The average magnitude of
$f_{i}$ is used to assess the divergence free.
Where $A_{i}$ is the surface area of the differential volume ($\bigtriangleup\textbf{V}_{i}$).
$\sigma_{J}$ and $\langle|f_{i}|\rangle$ should be equal to zero, when
the force- and divergence-freeness of extrapolated field are fully
satisfied. Fig\ref{Fig01} gives the distributions of $\sigma_{J}$ and $\langle|f_{i}|\rangle$ for these 23 active regions.
And for each criterion, the corresponding mean values are given in this figure.
It can be found that the $\sigma_{J}$ ranges from 0.21 (Rad) to 0.76 (Rad) with mean values 0.40 (Rad),
while $\langle|f_{i}|\rangle$ ranges from 0.13$\times 10^{-4}$ to 0.83$\times 10^{-4}$ with mean values 0.49$\times 10^{-4}$.
The amplitudes of $\sigma_{J}$ and $\langle|f_{i}|\rangle$ are comparable to the previous studies \citep{liu11a, whe00}.
Another aspect to assess the quality of extrapolated field can be done by comparisons of magnetic field lines with coronal images.
To see the consistences between field lines and coronal loops, here three examples are given in Fig\ref{Fig0-1}, \ref{Fig0-2} and \ref{Fig0-3}, respectively.
Here the left panel shows the magnetic field lines which are over plotted on the EIT/SOHO image, and the pure coronal image
are given in right panel to see the coronal image clearly.
It can be seen that the magnetic field lines can match coronal loops for global structures at some extent, especially for closed field lines
(red lines plotted in left panels). On the whole the extrapolated field for these active regions approximately satisfy conditions of force-free equations since the
criterion of force-freeness and divergence-freeness are acceptable at some extent, and the extrapolated field lines can match coronal loops for global structures. So it is reasonable to tentatively study the properties of spatial magnetic fields for these active regions using the extrapolated fields.

\begin{figure}
\epsscale{1.0} \plotone{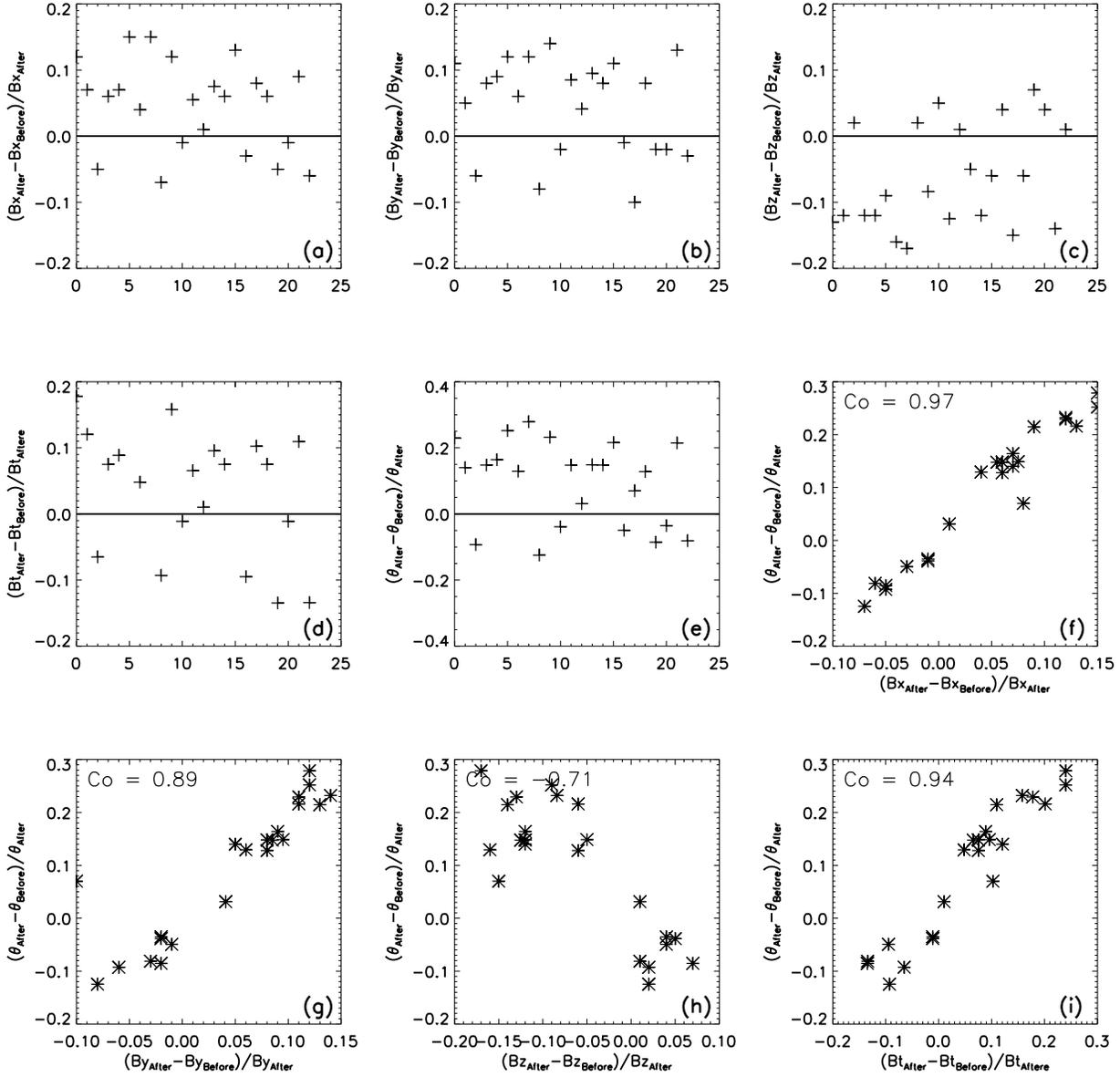} \caption{Plots of a, b, c, d and e show the changes of magnetic components of $B_{x}$, $B_{y}$, $B_{z}$, $B_{t}$ and the inclination angle ($\theta$), respectively. For each component, the differences of those component between after and before flare are normalized to its values after flare.
Here inclinations angles are calculated and averaged basing on closed field lines.
Plots of f, g, h and i show the correlations between the changes of $B_{x}$, $B_{y}$, $B_{z}$, $B_{t}$ and the changes of inclination angle ($\theta$),
and the correlation coefficients are calculated and labeled on corresponding panels.}\label{Fig1}
\end{figure}

Fig\ref{Fig1} shows the changes amplitudes of magnetic components of $B_{x}$, $B_{y}$, $B_{z}$, $B_{t}$ and the inclination angle,
which is defined by $atan(B_{t}/B_{z})$, $B_{t}$ is the transverse magnetic field ($B_{t}=\sqrt{B_{x}^{2}+B_{y}^{2}}$).
For each component, its differences between corresponding values after and before flare are normalized to its values after flare.
So the positive/negative values mean the increases/decreases of this component after the occurrences of flares,
and the values indicate the relative rates of increases after solar flares.
As for the inclinations angles, here only closed field lines are calculated and averaged
to avoid the biased that may originate from open field lines.
For the value of each individual component, it is calculated from the magnetic fields amplitudes of the whole space of active region. For example,
$B_{x}$=(1/N)$\sum <|B_{x}(i,j,k)|>$ (i, j and k is the pixel position of 3D space,
N is  the total number of vectors in the volume to be calculated) is the average of absolute values of magnetic fields.
In this figure, the plots of a, b, c, d and e correspond to the components of
$B_{x}$, $B_{y}$, $B_{z}$, $B_{t}$ and the inclination angle, respectively.
The changes of those magnetic components can indicate the topology changes of space magnetic fields, hence they are chosen to be studied firstly.
For example, the increases of transverse magnetic field can mean that magnetic
field lines become more lower after flare than those before solar flare. Also the increases/decreases of inclination angles indicate
most evidently that magnetic field lines become lower/higher after solar flare.
In figure 1, it can be found that the number of active with the increases of $B_{x}$, $B_{y}$, $B_{z}$, $B_{t}$ and inclination angle are
16, 15, 8, 16 and 16, respectively. It is found that the change amplitudes of $B_{x}$, $B_{y}$, $B_{z}$, $B_{t}$ do not exceed
20\%, but the change amplitudes of inclination angle exceed 20\% for a few active regions.
Through the distributions of inclination angles it can be found that there are 69\% (16/23) active region with
field line lower after flare. In this figure, the plot of f, g, h and i shows the correlations between
inclination angle and $B_{x}$, $B_{y}$, $B_{z}$, $B_{t}$. It can be found that there are positive
correlations between the inclination angle and $B_{x}$, $B_{y}$, $B_{t}$,
the correlation coefficients are 97\%, 89\% and 94\% between the inclination angle and $B_{x}$, $B_{y}$, $B_{t}$, respectively. There is weak negative correlations between the inclination angles and $B_{z}$ with correlation coefficient of 71\%.

\begin{figure}
\epsscale{1.0} \plotone{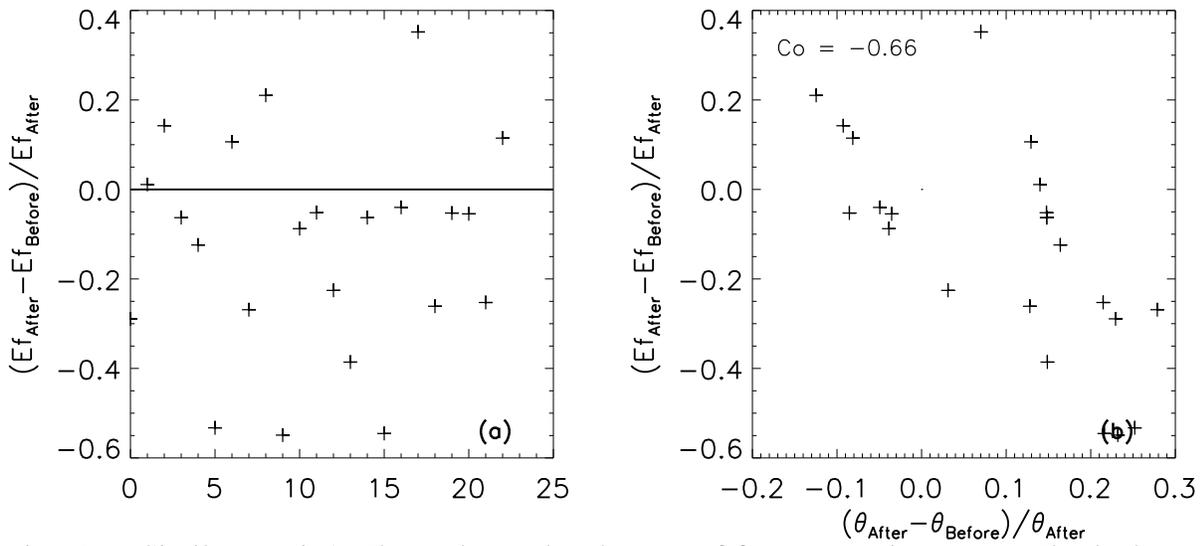} \caption{Similar as Fig1, Plot a shows the changes of free magnetic energy. Plot b shows the correlations between the changes of free energy and the changes of inclination angle ($\theta$), and the correlation coefficient is calculated and labeled. }\label{Fig2}
\end{figure}

It is considered traditionally that free magnetic energy is more related to solar flare.
The free magnetic energy is the energy difference between actual magnetic field and
the corresponding potential field (the minimum energy state
for a given photosphere boundary condition). In the practical application,
free magnetic energy can be calculated by the following formula:
\begin{equation}
E_{free}=E_{N}-E_{P}=\int\dfrac{B_{N}^{2}}{2\mu}dV-\int\dfrac{B_{P}^{2}}{2\mu}dV
\end{equation}
where V is the computational volume from photosphere to corona, and the superscripts N and P represent the
non-linear force free field (approximate actual space magnetic field of active region) and the potential field, respectively.
Fig\ref{Fig2} shows the changes of free energy of those active regions. Same as Fig\ref{Fig1}, the differences of free energy
between after and before flare are normalized to its values after flare.
Also, the positive/negative values mean the increase/decrease of free energy of corresponding active region.
From plot a, it can be found that there are 74\% (17/23) active regions with free energy decrease after solar flare.
In plot b, the correlations between changes of free energy and inclination angle are shown, it can be found that there
exist weak negative correlations with correlation coefficient of 66\%. Additionally, for 6 active regions with free energy increase, there are 3 active regions with inclination decrease. While for 17 active regions with free energy decrease, there are only 4 active regions with inclination angle decrease.

\section{DISCUSSIONS AND CONCLUSIONS}
In this work, the statistical researches on the properties of space magnetic field
before and after solar flare are presented basing on
observations obtained by SMFT installed at HSOS.
The regularization of magnetic topology indicated partially by the changes of inclination angle
can display the evolutions of active region spatial magnetic fields.
While free magnetic energy is related the most profound physical essence, which can explicate the
productions of solar flare undoubtedly, due to solar flare energy come from free magnetic energy.
So the main studies are focused on the changes of inclination angles and free magnetic energy
before and after solar flare in this research.

The selection criteria described in the above section for active regions with flare eruption
are very rigorous in this study, since the precise changes of magnetic field after solar flare
are the primary focuses, at last 23 active regions are chosen as data sample in this study.
Through calculations, it is found that 16 (69\%) active regions with inclination angle increases, it is means that
the transverse fields increase relatively after solar flare, which can match the previous
theoretical model \citep{hu08, hu11}.
For free magnetic energy, there are 17 (74\%) active regions with the decreases of free energy, which
are similar to some previous research results \citep{sun12}.
It is noted for the changes of free magnetic energy that changes amplitudes of free energy decreases are
larger than those of free energy increases. Similarly, the changes amplitudes of inclination angle increases are
larger than those of inclination angle decreases.

In this paper, the sensitive magnetic parameters related to the productions of solar flare are calculated
from the observations. It can not give the exact conclusions which parameter is direct related to solar
flare. However, it can be concluded that the results of most active regions agree with the previous studies,
which have given the conclusions that the transverse magnetic field (free magnetic energy) will increase (decrease)
after the eruption of solar flare.
For active regions not match the classical theoretical model, there maybe some more complicated factors that
lead to the corresponding results.
Additionally, it is should noted that there are some uncertainties on observation data and numerical extrapolation,
which all can affect the precise of statistical results.

\begin{table}
\begin{center}
\caption{The active regions selected.}
%\begin{tabular}{cccccccccccc}
\begin{longtable}{cccccccccccc}
\tableline\tableline
NOAA      &Date            &Observation Time       &Flare Time              &Observation Time   &Flare       &Flux Deviation\\
          &                &Before Flare           &(UT)                    &After Flare (UT)   &Level       &$(F_{p}-F_{n})/F_{p}$ \\
          &                &                       &                        &                   &            &Before|After \\
\tableline
08040     &1997-05-21      &04:13:21                &06:04                  &06:18:21           &C$_{2.7}$         &12\%|15\% \\

08156     &1998-02-15      &03:26:03                &04:20                  &04:58:49           &C$_{1.8}$         &15\%|16\% \\
08218     &1998-05-12      &00:21:58                &04:49                  &07:07:51           &C$_{1.2}$         &17\%|16\% \\
08319     &1998-08-28      &23:07:55                &00:36                  &01:48:13           &C$_{1.2}$         &16\%|14\% \\
08369     &1998-10-28      &01:43:09                &02:03                  &02:21:44           &C$_{1.6}$         &13\%|9\% \\

08716     &1999-10-07      &04:21:36                &05:19                  &06:32:08           &C$_{2.2}$         &12\%|10\% \\
08728     &1999-10-16      &04:32:18                &05:05                  &06:31:26           &C$_{2.3}$         &8\%|11\% \\
08737     &1999-10-21      &04:31:49                &06:20                  &06:40:21           &C$_{1.9}$         &15\%|12\% \\
08925     &2000-03-30      &04:32:18                &05:05                  &06:31:26           &C$_{2.3}$         &11\%|10\% \\

09054     &2000-06-16      &05:16:59                &06:31                  &07:03:40           &C$_{3.2}$         &17\%|12\% \\
09154     &2000-09-06      &02:25:45                &03:38                  &04:04:07           &C$_{1.7}$         &9\%|14\% \\
09321     &2001-06-16      &05:40:01                &07:08                  &07:52:06           &C$_{1.0}$         &12\%|10\% \\
09463     &2001-05-25      &23:02:45                &23:57                  &03:28:32           &C$_{1.7}$         &10\%|11\% \\
09531     &2001-07-11      &00:20:17                &00:41                  &01:34:35           &C$_{1.0}$         &13\%|15\% \\

10061     &2002-08-09      &04:37:43                &06:36                  &06:48:31          &C$_{1.4}$         &12\%|9\% \\
10119     &2002-09-20      &01:07:31                &01:56                  &03:55:21          &C$_{2.4}$         &6\%|8\% \\
10137     &2002-10-03      &01:25:53                &02:15                  &03:00:06          &M$_{2.1}$         &7\%|9\% \\

10488     &2003-10-27      &23:50:30                &00:41                  &01:14:24          &C$_{5.3}$         &9\%|11\% \\

10767     &2005-05-27      &04:31:00                &05:00                  &05:25:38          &C$_{2.5}$         &13\%|11\% \\
10786     &2005-07-05      &01:11:46                &01:24                  &01:48:02          &M$_{1.8}$         &14\%|11\% \\
10826     &2005-12-03      &05:06:01                &05:31                  &06:06:42          &M$_{3.6}$         &12\%|13\% \\

10953     &2007-5-02      &23:27:54                 &23:28                  &00:03:26          &C$_{8.5}$         &14\%|12\% \\
10960     &2007-06-05      &03:48:36                &04:15                  &04:31:49          &C$_{1.2}$         &15\%|17\% \\

\tableline
%\end{tabular}
\end{longtable}
%% Any table notes must follow the \end{tabular} command.
%\tablenotetext{a}{refers to the maximum amplitude of intensity oscillations.}
\end{center}
\end{table}

\acknowledgments
%The authors thank the anonymous referee for helpful comments and suggestions.
This work was partly supported by the Grants: 2011CB811401, KLCX2-YW-T04, KJCX2-EW-T07,
11203036, 11221063, 11178005, 11003025, 11103037, 11103038, 10673016, 10778723 and 11178016, the
Young Researcher Grant of National Astronomical Observations,
Chinese Academy of Sciences, and the Key Laboratory of Solar
Activity National Astronomical Observations, Chinese Academy
of Sciences.


\clearpage

%% the file, uncomment the \plotone, \plottwo, and
%% If you need a layout that cannot be achieved with \plotone or
%% \plottwo, you can invoke the graphicx package directly with the
%% \includegraphics command or use \plotfiddle. For more information,
%% please see the tutorial on "Using Electronic Art with AASTeX" in the
%% documentation section at the AASTeX Web site,
%% http://www.journals.uchicago.edu/AAS/AASTeX.
%%
%% The examples below also include sample markup for submission of
%% supplemental electronic materials. As always, be sure to check
%% the instructions to authors for the journal you are submitting to
%% for specific submissions guidelines as they vary from

%% This example uses \plotone to include an EPS file scaled to
%% 80% of its natural size with \epsscale. Its caption
%% has been written to indicate that additional figure parts will be

\clearpage


\clearpage

%% Here we use \plottwo to present two versions of the same figure,
%% one in black and white for print the other in RGB color
%% for online presentation. Note that the caption indicates
%% that a color version of the figure will be available online.
%%

\clearpage

\end{document}